\documentclass[prb,twocolumn,showpacs,preprintnumbers,amsmath,amssymb]{revtex4}

\usepackage{graphicx}
\usepackage{dcolumn}
\usepackage{bm}
\usepackage[pdftex, bookmarks={false}, pdfauthor={Rudro Rana Biswas}, pdftitle={Power laws in surface state LDOS oscillations near a step edge}]{hyperref}
\hypersetup{colorlinks=false, linkcolor=red, citecolor=green, filecolor=blue, urlcolor=blue, filebordercolor={.8 .8 1}, urlbordercolor={.8 .8 0}}
\usepackage[all]{hypcap}

\newcommand\ba{\begin{array}}
\newcommand\ea{\end{array}}
\newcommand\nn{\nonumber}
\newcommand\ri{\right}
\renewcommand\le{\left}

\newcommand{\feyn}[1]{#1\kern-0.45em/}

\newcommand{\tto}{\rightarrow}

\renewcommand\a{\alpha}

\renewcommand\b{\beta}

\renewcommand\c{\psi}

\renewcommand\d{\delta}
\newcommand\D{\Delta}

\newcommand\f{\phi}

\newcommand\g{\gamma}

\newcommand\G{\Gamma}

\newcommand\mbk{\mbs{k}}

\newcommand\m{\mu}

\newcommand\rr{\rho}

\newcommand\vx{\chi}
\newcommand\mbx{\mbs{x}}

\newcommand\y{\eta}

\newcommand\mbs{\boldsymbol}

\begin{document}
\title{Power laws in surface state LDOS oscillations near a step edge}
\author{Rudro R. Biswas$^{1,3}$}
\email{rrbiswas@physics.harvard.edu}
\author{Alexander V. Balatsky$^{2,3}$}%
\affiliation{%
$^{1}$Department of Physics, Harvard University, Cambridge, MA 02138\\
$^{2}$Theoretical Division, Los Alamos National Laboratory, Los Alamos, NM 87545\\
$^{3}$Center for Integrated Nanotechnologies, Los Alamos National Laboratory, Los Alamos, NM 87545
}
\date{\today}
\begin{abstract}
In this paper we indicate a general method to calculate the power law that governs how electronic LDOS oscillations decay far away from a surface step edge (or any local linear barrier), in the energy range when only 2D surface states are relevant. We identify the critical aspects of the 2D surface state band structure that contribute to these decaying oscillations and illustrate our derived formula with actual examples.
\end{abstract}
\pacs{03.65.Nk, 07.79.Cz, 73.20.-r, 73.20.At}
\maketitle

\section{Introduction}

For over a couple of decades now Scanning-Tunneling Microscopy (STM) experiments have been used to observe the effects of perturbations to electronic surface states, in the form of atomic defects, corrals and step edges\cite{1993-crommie-vn,1997-yazdani-kx,2010-alpichshev-fk,2010-richardella-uq}. The quantum electronic response to defects can give us basic information about the band structure of the scattered quasiparticles\cite{1998-petersen-fk}; it also encodes information about their nature and this is useful when probing correlated phases\cite{1997-yazdani-kx,2002-hoffman-kx,2002-hoffman-vn,2000-pan-uq}. In this paper, we shall consider the case of a step edge on a 2D surface and calculate the spatial decay of standing waves created in the surface LDOS, far away from the step edge. We shall show, by a simple process of power counting, that the geometry of the constant energy cut of the quasiparticle band structure and a qualitative knowledge of the character of the quasiparticle wavefunctions provide enough information to pin down the power with which the LDOS oscillations decay far away from the step edge.

\section{Theory}
\begin{figure}[b]
\begin{center}
\resizebox{9.5cm}{!}{\includegraphics{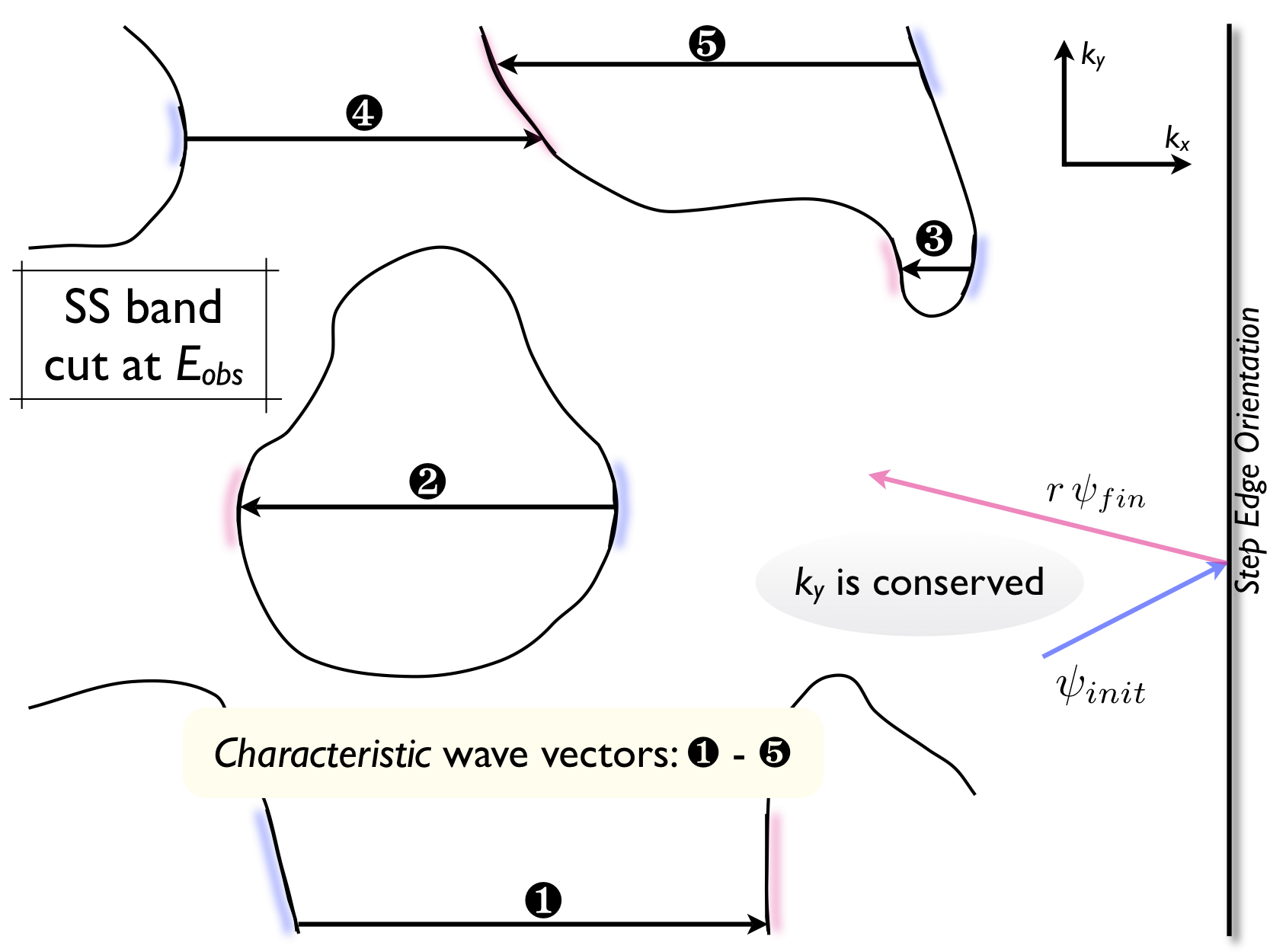}}
\caption{(Color online) A constant energy cut of a generic 2D electronic surface state (SS) band structure, taken at the energy $E_{\text{obs}}$ of an STM probe; on the right is shown the orientation of the surface step edge (or any linear barrier) in question. Quasiparticles are scattered `horizontally', preserving $k_{y}$. The regions where the scattering wave-vectors vary the slowest (locally) with $k_{y}$ are shaded -- blue for incident and pink for reflected states. The `characteristic' scattering wave vectors are also indicated by arrows and numbers.}
\label{fig-bandcut}
\end{center}
\end{figure}

We begin by considering a general band structure for the 2D surface states, whose cross-section at the energy of observation $E_{\text{obs}}$ is shown in Figure \ref{fig-bandcut}. Also shown in the figure is the orientation of the surface step edge --- parallel to the $y$-direction. Because of the conservation of momentum parallel to the edge during a scattering process, the incoming and outgoing states must be connected by straight lines \emph{perpendicular} to the step edge. Some such processes are also marked in Figure \ref{fig-bandcut}. The arrows joining the initial and final states denote the wave-vector of LDOS oscillations that particular scattering process would give rise to. Obtaining the total LDOS involves summing up these oscillations. The most \emph{coherent} contributions to this sum come from regions where the scattering wave-vectors change the slowest as we move parallel to the step edge, i.e, changing only the $k_{y}$ of the scattering states. We denote the `identifying' scattering wave vector in each such region as the `characteristic' wave vector of that region.

As a very common example, for a circular constant energy cut as in a 2DEG (Figure \ref{fig-2deg}), the most coherent contributions come from the scatterings around the diameter --- the characteristic wave vector in this case is thus the diameter, $\D_{0}$.

The new electronic LDOS far away from the step edge is now provided by (below, `new' refers to the new energy eigenstates while `init' and `fin' refer to the initial and final scattering states, respectively):
\begin{align}
\rr(x,E) &= \sum_E \le|\c_{\text{new}}\ri|^2\nn\\
&=\sum_E \le|\c_{\text{init}} + r \c_{\text{fin}}\ri|^2 + \text{transmitted from other side}\nn\\
&= \overbrace{\sum_E \le(\le|\c_{\text{init}}\ri|^2 + \le|r \c_{\text{fin}}\ri|^2 + \text{transmitted}\ri)}^{x-\text{independent part}} \nn\\
& \qquad \qquad \qquad + 2 \sum_E\text{Re}\le[r\; \c^\dag_{\text{init}}\cdot\c_{\text{fin}}\ri]
\end{align}
Writing the energy-momentum eigenstates as $\c_{\mbk}(\mbx) = \vx e^{i\mbk\cdot\mbx}$, where $\vx$ denotes an `internal' part involving the spin and other internal components, the $x$-dependent part of the LDOS can be summarized as
\begin{align}\label{eq-master}
\d\rr(x,E) &\propto \sum \int_{0} dk_{y} \;\rr_{0}(k_{y}) \text{Re}\le[r(k_{y})\le(\vx_{f}^{\dag}\cdot\vx_{i}\ri)e^{i\D k_{x} x}\ri]
\end{align}
The sum is over the various regions of coherent scattering, each one corresponding to a characteristic scattering vector. The outer limits of these integrals are not important as the oscillations there de-cohere rapidly. $\rr_{0}(k_{y})$ is a DOS factor (it multiplicatively converts the measure $dk_{y}$ to a product of the \emph{length} of the band curve enclosed between $k_{y}$ and $k_{y}+dk_{y}$ and the DOS in that region).

The $x$-dependence of a characteristic oscillation far away from the step edge may be found from the above expression by writing down the lowest order $k_{y}$-dependencies of the relevant quantities near each characteristic wave-vector ($\d k_{y}$ is the $k_{y}$-displacement from the associated characteristic wave vector):
\begin{align}\label{eq-powers}
\rr_{0}(k_{y})\sim \rr_{0}\d k_{y}^{\a}\nn\\
r(k_{y}) \sim r_{0}\d k_{y}^{\b}\nn\\
\vx_{f}^{\dag}\cdot\vx_{i} \sim \varrho \d k_{y}^{\g}\nn\\
\D k_{x}\sim \D_{0} + \D_{1}\d k_{y}^{\y}
\end{align}
Changing the integration variable $\d k_{y}$ to the variable $\m = \d k_{y}^{\y} \,x$ in \eqref{eq-master} and using \eqref{eq-powers}, we obtain our central result
\begin{align}\label{eq-finalresult}
\d\rr(x,E) &\propto \sum \frac{\rr_{0}}{x^{\frac{\a+\b+\g+1}{\y}}}\int \frac{d\m}{\m} \;\m^{\frac{\a+\b+\g+1}{\y}} \text{Re}\le[r_{0} \varrho \,e^{i(\D_{0} x + \D_{1}\m)}\ri]\nn\\
&\sim \sum \le|\rr_{0} r_{0} \varrho \ri| \frac{\sin(\D_{0} x + \f)}{x^{(\a+\b+\g+1)/\y}}
\end{align}
This asymptotic behavior is correct for $x\gg (\D k)^{-1}$, where $\D k$ is the characteristic size of the region in momentum space where the scaling laws \eqref{eq-powers} hold. The power of decay of the oscillations coming from each characteristic region is thus given by $(\a+\b+\g+1)/\y$, which may be evaluated using our knowledge of the band structure in that region. Note, however, that we haven't been able to evaluate the total strength of the scattering process which requires a much more detailed calculation including evaluating the reflection amplitudes themselves. This exercise provides us with the combination of possible power laws we can try to fit actual experimental data to, if an idea of the band structure exists.

\section{Examples}
\subsection{2DEG\cite{1993-crommie-vn}}
\begin{figure}[h]
\begin{center}
\resizebox{6cm}{!}{\includegraphics{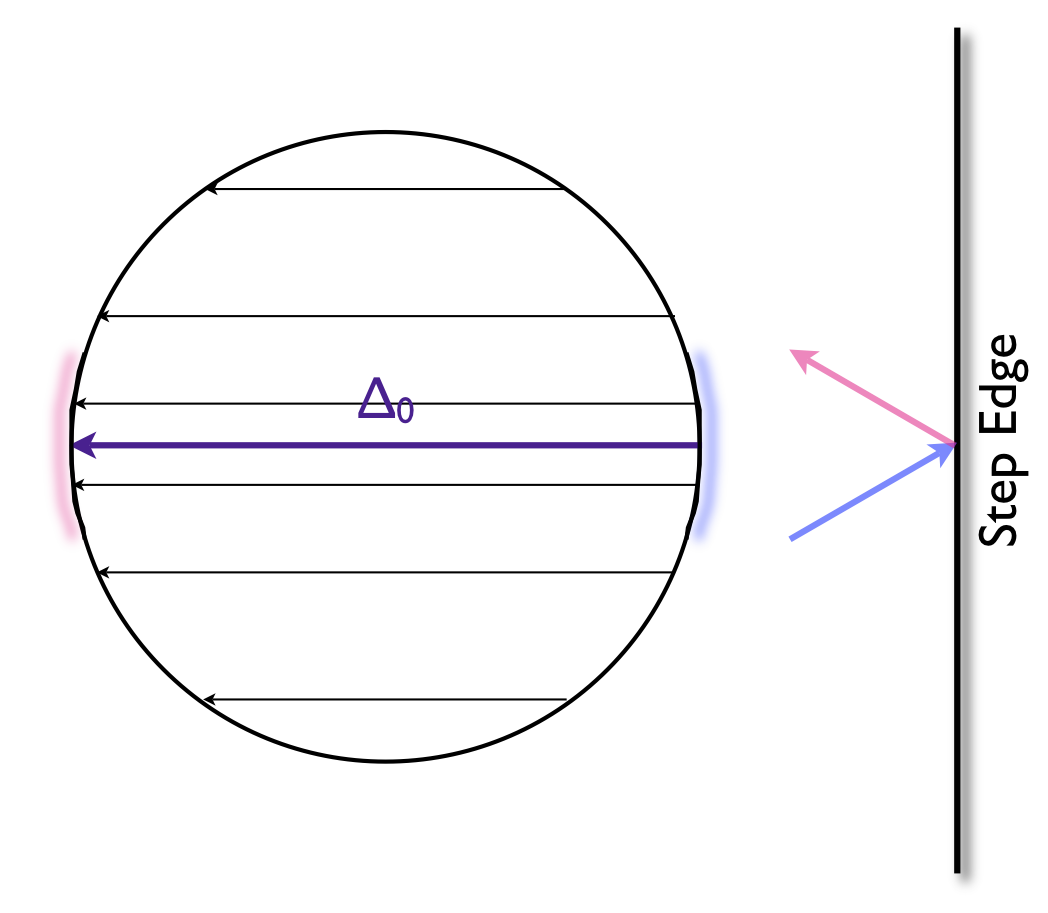}}
\caption{(Color online) The characteristic wave-vector $\D_{0}$ in the case of a circular band (for 2DEGs with rotational invariance)}
\label{fig-2deg}
\end{center}
\end{figure}

The scattering wave-vector varies slowest around the equator (see Figure \ref{fig-2deg}). Thus, $\D_{0}$ = diameter of circle $= 2k_{E}$, $\a=\b=\g=0$ (assuming, quite reasonably, that the reflection amplitude is nonzero and smooth across normal scattering). Also, from the geometry of the band, we get $\y=2$. Using \eqref{eq-finalresult}, this tells us that:
\begin{align}
\d\rr(x,E) \sim \frac{\sin(\D_{0} x + \f)}{x^{1/2}} \qquad (\D_{0} = 2k_{E})
\end{align}

\subsection{Strong Topological Insulator (circular band cut)}
\subsubsection{Generic barrier}
\begin{figure}[h]
\begin{center}
\resizebox{6cm}{!}{\includegraphics{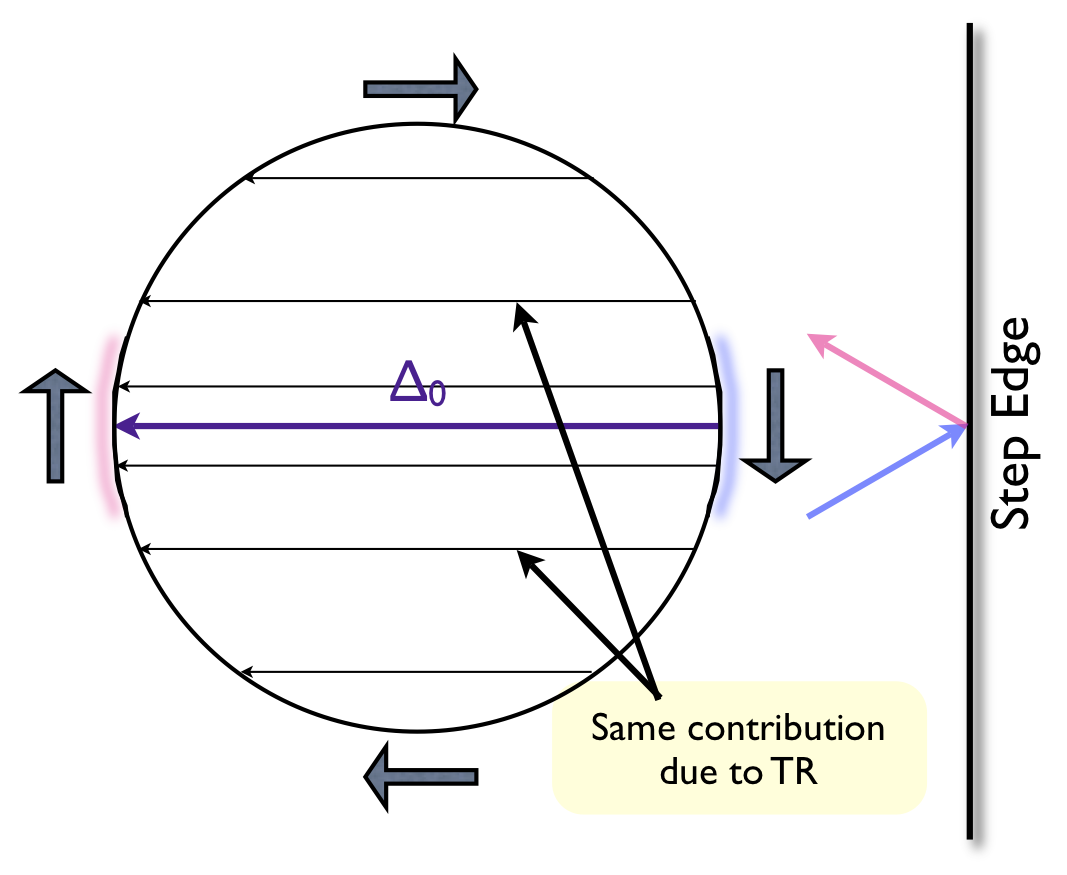}}
\caption{(Color online) Constant energy cut of a circular surface band on a STI surface. The spins, indicated as block arrows, are antiparallel for normal scattering (the characteristic scattering process for the STI surface state band that is circular) -- the spin overlap magnitude is thus generically a linear function of the angle of incidence. The same may be said for the reflection amplitude magnitude which is a certain gauge is antisymmetric in the angle of incidence.}
\label{fig-stidiam}
\end{center}
\end{figure}

This case is illustrated in Figure \ref{fig-stidiam} and is realized for the gapless surface states in Strong Topological Insulators like Bi$_{2}$Se$_{3}$ and Bi$_{2}$Te$_{3}$ (at energies near the Dirac point). The scattering wave-vector varies slowest around the equator (as in the 2DEG case), where $\D_{0}$=diameter of circle, $\a=0$, $\b=1$ since the reflection coefficient changes sign \cite{2009-biswas-fk} as one crosses the diameter/case of normal reflection (can be any odd power; should be linear \emph{generically}), $\g=1$ because the spins are exactly antiparallel for scattering states at the diameter and thus the lowest order overlap is linear in $\d k_{y}$, and $\y=2$ as in the 2DEG case. This gives rise to:
\begin{align}\label{eq-stigeneric}
\d\rr(x,E) \sim \frac{\sin(\D_{0} x + \f)}{x^{3/2}} \qquad (\D_{0} = 2k_{E})
\end{align}
This result agrees with numerical calculations for particular cases of the model describing the step edge\cite{2009-zhou-fk}.

\subsubsection{`Perfect' reflection}
The scattering wave-vector varies slowest around the equator (as in the 2DEG case), where $\D_{0}$=diameter of circle, $\a=\b=0$ \cite{2009-biswas-fk} (since the reflection amplitude is constant in magnitude near normal incidence), $\g=1$ as argued for the previous case, and $\y=2$. This gives rise to:
\begin{align}
\d\rr(x,E) \sim \frac{\sin(\D_{0} x + \f)}{x} \qquad (\D_{0} = 2k_{E})
\end{align}

In the actual case, there will always be a region near normal incidence where the reflection amplitude will become linear (because it is antisymmetric). This means that `very' far away $\sim$ the inverse of the $k_{y}$-span of the region where $r$ is linear, the previous result \eqref{eq-stigeneric} for the generic barrier should hold.

\subsection{Bi$_{2}$Te$_{3}$ (with hexagonal warping)\cite{2010-alpichshev-fk}}

If we are far away from the Dirac point, the surface band of Bi$_{2}$Te$_{3}$ exhibits hexagonal warping\cite{2009-fu-fk}. The following results hold when the STM bias maintains our observation energy in that regime.
\subsubsection{Step edge $\perp$  $\G$M direction}
\begin{figure}[h]
\begin{center}
\resizebox{6cm}{!}{\includegraphics{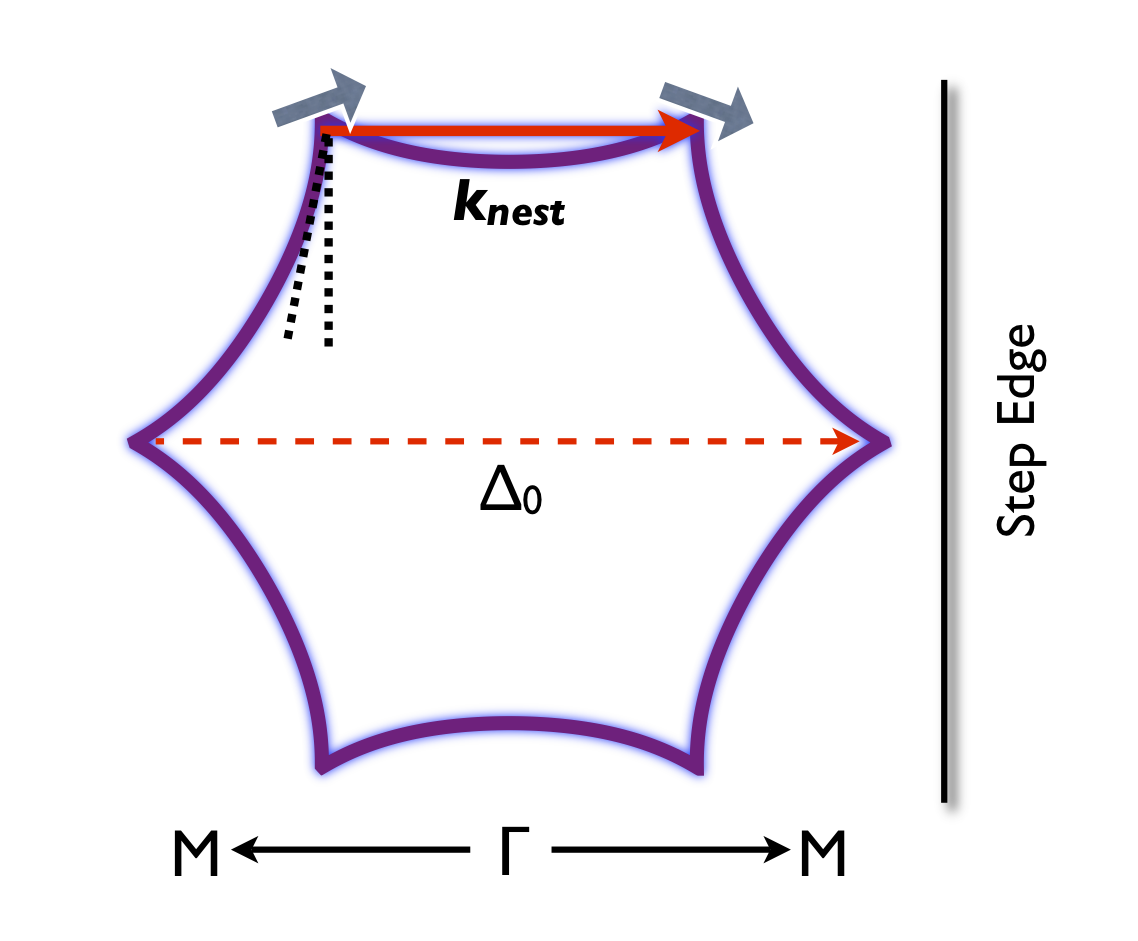}}
\caption{(Color online) Scattering processes from a step edge on Bi$_{2}$Te$_{3}$ oriented perpendicular to the $\G$M direction, in the energy range where the band exhibits hexagonal warping. The scattering vector varies linearly near $k_{\text{nest}}$, as indicated by the angle made by the dotted lines, leading to $\y=1$. The weaker characteristic scattering process is denoted by the dotted arrow. Spins are indicated as block arrows.}
\label{fig-linearband}
\end{center}
\end{figure}

The scattering wave-vector varies slowest around the equator (as in the 2DEG case), where $\D_{0}$=diameter of circle, $\a=0$, $\b=1$, $\g=1$ and $\y=2$ exactly as argued before for the circular STI band. However, the extent of this region is very small and the scattering is found to be \emph{dominated} by processes connecting the hexagonal `corners' (marked by bold arrow in Figure \ref{fig-linearband}), with a characteristic scattering vector $k_{\text{nest}}$ \cite{2010-alpichshev-fk}. For the latter case, since the spin states have a finite overlap with each other at the hexagon corners\cite{2009-fu-fk}, we have $\g=0$. Also, assuming that the reflection coefficient is smooth for the relevant scattering processes, $\b=0$. Finally, $\a=0$ (DOS is finite and smooth) and the overwhelmingly `linear' nature of the bands yield $\y=1$. Putting these together, we obtain the observed variation\cite{2010-alpichshev-fk}
\begin{align}
\d\rr(x,E) \sim \frac{\sin( k_{\text{nest}}x + \f)}{x}
\end{align}

\subsubsection{Step edge $\perp$  $\G$K direction}
The wave-vectors vary slowest around the equator (as in the 2DEG case) and from the considerations of the cicular STI surface band above, we can conclude that there should be characteristic oscillations at $2k_{\G K}$ decaying as $x^{-3/2}$ (or $1/x$ for a `perfect' reflector). In this case, because of the larger extent of the characteristic scattering region around the diameter, these oscillations may be strong and observable. Because of the presence of the linear band shape with larger spectral presence and reflection strengths near the corners, we can also observe LDOS oscillations from the corner$\tto$corner scattering processes, decaying as $1/x$. Of course, from our simple calculation we cannot reliably predict which of the two processes discussed above have the stronger signature.
\begin{figure}[h]
\begin{center}
\resizebox{6cm}{!}{\includegraphics{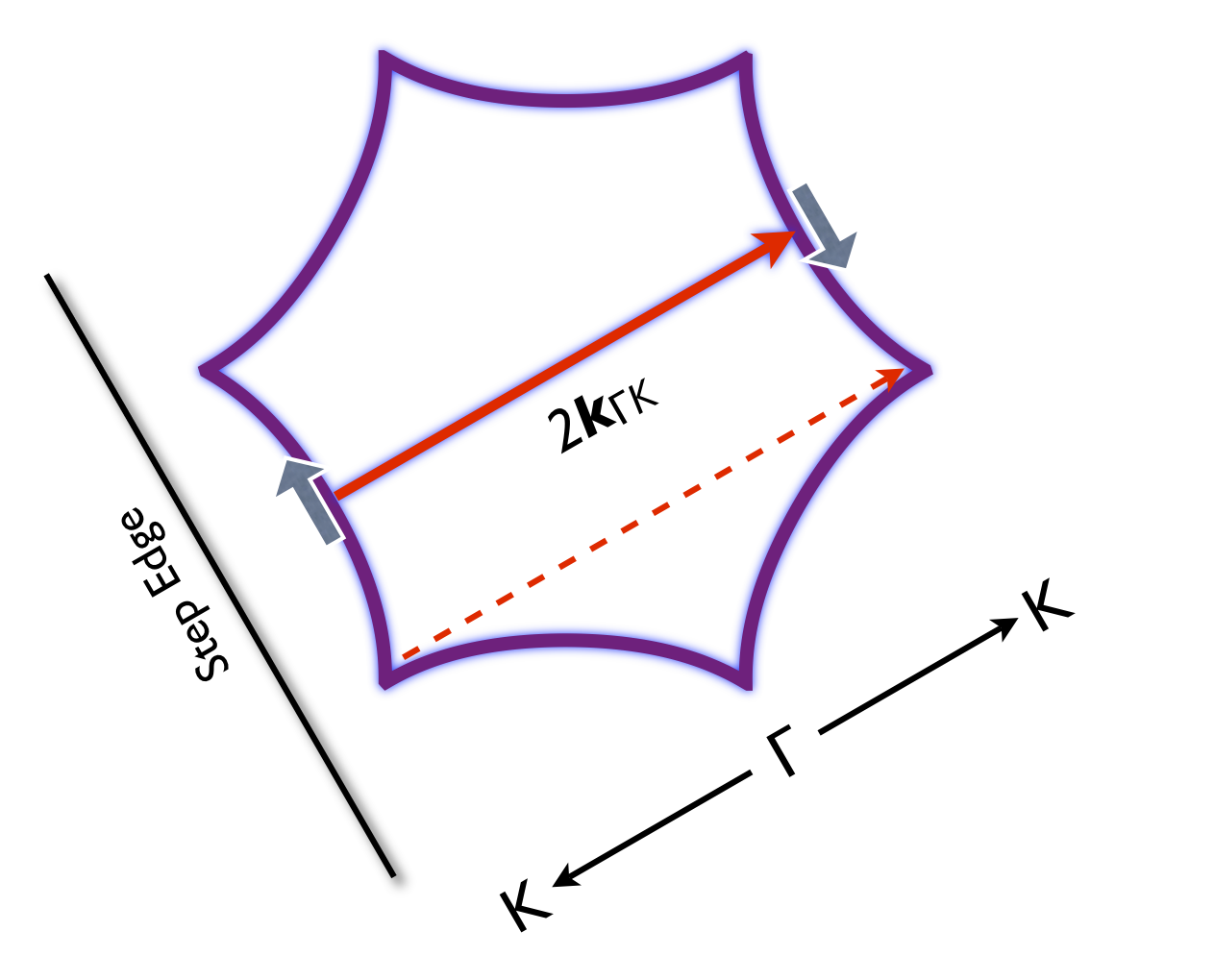}}
\caption{(Color online) Scattering processes from a step edge on Bi$_{2}$Te$_{3}$ oriented perpendicular to the $\G$K direction, in the energy range where the band exhibits hexagonal warping. Spins are indicated as block arrows.}
\label{fig-bi2se3}
\end{center}
\end{figure}

\section{Isolating contributions using the 1-D Fourier Transform}
The Fourier Transform of the LDOS data may be used to observe signatures from more than one set of scattering processes. For oscillations decaying as $\sin(K x + \f)/x^{n}$, scaling analysis tells us that the Fourier transform looks like $F(k) \sim |k\mp K|^{n - 1}$, when $k\sim\pm K$. Thus, one way to look for contributions to these oscillations would be to scan the 1-D Fourier transform of the LDOS (taken over the long-distance behavior) for features at the `characteristic' scattering vectors. The FT near those points can then be fitted to the abovementioned power laws (or a logarithm, for the case of a $1/x$ decay) to recover the spatial decay power laws.

\section{Conclusion}
We have outlined a method to calculate the possible oscillatory power laws governing the decay of LDOS perturbations next to a step edge or some such linear barrier on a surface, in the energy range when only electronic surface states are relevant and the surface band structure is qualitatively known. To find these laws we need to identify the characteristic scattering regions (Figure \ref{fig-bandcut}), compute the scaling powers of the relevant quantities \eqref{eq-powers} and from that obtain the possible oscillatory powers \eqref{eq-finalresult}.

\acknowledgements
This work was supported by the US DOE thorough BES and LDRD and by the University of California UCOP program T027-09. We would also like to acknowledge illuminating discussions with M.\ Crommie, D.\ Haldane, H.\ Manoharan and members of M.\ Z.\ Hasan and A.\ Yazdani's research groups.

\end{document}